\documentclass{PoS}

\title{On high energy exclusive leptoproduction of the $\rho$ meson up to twist 3}

\ShortTitle{Exclusive leptoproduction of the $\rho$ meson up to twist 3}

\author{I. V. Anikin\\
 Bogoliubov Laboratory of Theoretical Physics, JINR,
             141980 Dubna, Russia}
\author{\speaker{A. Besse}\\
       LPT, Universit\'e Paris-Sud, CNRS, 91405 Orsay, France\\
      E-mail: \email{besse@th.u-psud.fr}
}
\author{D.Yu. Ivanov\\
Sobolev Institute of Mathematics and Novosibirsk State University, 630090 Novosibirsk, Russia}
 \author{B. Pire\\
 CPhT, \'Ecole Polytechnique, CNRS, 91128 Palaiseau, France}
\author{L. Szymanowski 
\\NCNR, Warsaw, Poland}
\author{S. Wallon\\
 LPT, Universit\'e Paris-Sud, CNRS, 91405 Orsay, France and UPMC Univ. Paris 06, facult\'e de physique, 4 place Jussieu, 75252 Paris Cedex 05, France}
\abstract{The Regge inspired $k_T$-factorization scheme  expresses the scattering amplitude for exclusive leptoproduction of a light vector meson $\rho$ in terms of two impact
factors, one for the $\gamma^* - \rho$ transition and the other one for the nucleon to nucleon transition,
with, at leading order, a two "reggeized" gluon exchange in the $t$-channel. We report here on a recent phenomenological study \cite{Anikin:2011sa} where we apply a previously developped scheme   \cite{Anikin:2009}  which
consistently takes into account the twist 3 distribution
amplitudes for transversely polarized $\rho$ meson.
The computation of the  $\gamma^* \to \rho$ impact factor are performed within collinear factorization of QCD, up to the twist 3 
level, based on the  Collinear Factorization scheme applied to the amplitudes $\gamma^{*}(\lambda_{\gamma})g(k) \rightarrow g(r-k)\rho(\lambda_{\rho})$.
 Our results are in good agreement with
high energy experimental data for the ratios of helicity amplitudes.
 }

\FullConference{The 2011 Europhysics Conference on High Energy Physics-HEP 2011,\\
		July 21-27, 2011\\
		Grenoble, Rh\^one-Alpes, France}

\begin{document}

In the impact factor representation at the Born order, the amplitude of the exclusive process $\gamma^{*}(q,\lambda_{\gamma}) \, N \rightarrow \rho(p_\rho,\lambda_{\rho}) \, N$ reads (using boldface letters for euclidean two-dimensional transverse vectors):
\begin{equation}
\label{defImpactRep}
 T_{\lambda_{\rho}\lambda_{\gamma}}(\textbf{r};Q , M) = is\int \frac{d^2\textbf{k}}{(2\pi)^2}\frac{1}{\textbf{k}^2(\textbf{k}-\textbf{r})^2} \Phi^{N \rightarrow N} (\textbf{k},\textbf{r};M^2)\Phi^{\gamma^{*}(\lambda_{\gamma}) \rightarrow \rho(\lambda_{\rho})}(\textbf{k},\textbf{r};Q^2)\,.
 \end{equation}
 The momenta $q$ and $p_\rho$ are parameterized via Sudakov decompositions in terms of two
 light cone vectors $p_1$ and $p_2$  as
$
q=p_1-\frac{Q^2}{s}p_2 ,  p_{\rho}=p_1+\frac{m_{\rho}^2-t+t_{min}}{s} p_2 + r_\perp\, ,  2\, p_1.p_2=s , 
$ where $Q^2=-q^2$ is the virtuality of the photon, and its presence justifies the use of perturbation theory, and $m_{\rho}$ is the mass of the $\rho$ meson. The impact factor
$ \Phi^{N \rightarrow N}$ in Eq.(\ref{defImpactRep}) cannot be computed within perturbation theory, and we use a simple  phenomenological model provided in Ref. \cite{Gunion:1976iy}, of the form:
\begin{equation}
\label{ProtonIF}
\Phi_{N \to N}(\textbf{k},\textbf{r};M^2)= A \, \delta_{ab}\left[\frac{1}{M^2+(\frac{\textbf{r}}{2})^2}-\frac{1}{M^2+(\textbf{k}-\frac{\textbf{r}}{2})^2}\right]\,.
\end{equation}
 $A$ and $M$  are free parameters that correspond to the soft scale of the proton-proton impact factor. This simple model can be interpreted  by assuming the existence inside the proton of some typical color-dipole configurations (onia) which will couple to the $\rho-$meson impact factor through a two-gluon exchange.  Such a model was  used successfully for describing DIS at small $x$ \cite{Navelet:1996jx}.

The calculation of the $\gamma^{*}(\lambda_{\gamma}) \rightarrow \rho(\lambda_{\rho})$ impact factor
$ \Phi^{\gamma^{*}(\lambda_{\gamma}) \rightarrow \rho(\lambda_{\rho})}$ is performed within collinear factorization of QCD. 
The
dominant contribution corresponds to the $\gamma^*_L \to \rho_L$ transition (twist 2), while the other transitions  are power suppressed.
The $\gamma^*_L \to \rho_L$ and  $\gamma^*_T \to \rho_L$ impact factors were computed long time ago \cite{Ginzburg:1985tp}, while a consistent treatment of the twist 3 
$\gamma^*_T \to \rho_T$ impact factor has been performed only recently in Ref.~\cite{Anikin:2009}.
It is based on the  collinear factorization  beyond the leading twist, applied to the amplitudes $\gamma^{*}(\lambda_{\gamma})g(k) \rightarrow g(r-k)\rho(\lambda_{\rho})$. These scattering amplitudes are the sum of the convolution of a hard part (denoted by $H$ and $H_\mu$ for two and three body contributions respectively), that corresponds to the transition of the virtual photon into the constituents of the $\rho$ meson and their interactions with off-shell gluons of the t-channel, and a soft part (denoted by $\Phi$ and $\Phi^\mu$). As the photon is highly virtual,
this convolution reduces to a factorized form, expressed as a convolution in the longitudinal momenta of  partons in the hard parts in collinear kinematics  with distribution amplitudes (DAs).

 Our accuracy is limited to dominant contributions both for $\gamma^*_L \to \rho_L$ (twist 2) and $\gamma^*_T \to \rho_T$ (twist 3) spin flip and nonflip transitions, therefore only leading terms of the expansion in $1/Q$ in both amplitudes are kept. Hence only two-body (quark-antiquark operators) and three-body (quark-antiquark gluon operators) non-local operators are involved.
 The chiral even 
$\rho$-meson DAs  up to twist 3 are defined by  matrix elements of non-local
light-cone operators. We perform a reduction of DAs to a minimal set $\varphi_1$, $B$, $D$,
thanks to QCD equations of motion and $n$-independency condition.

\begin{figure}[h!]
	\hspace{1cm}
\includegraphics[width=0.4\textwidth]{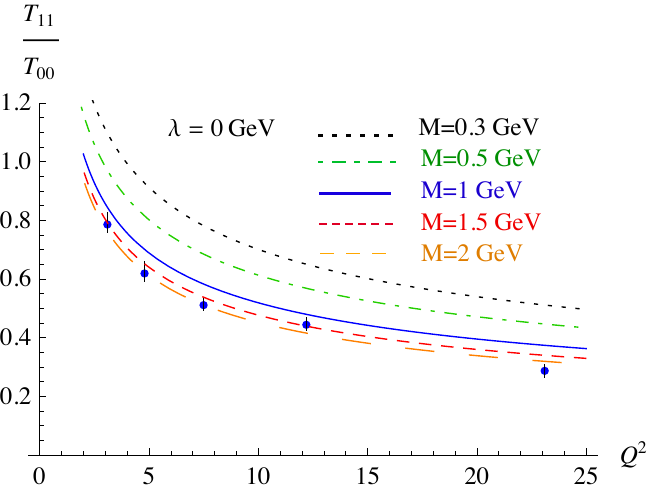} \hspace{1cm} \includegraphics[width=0.4\textwidth]{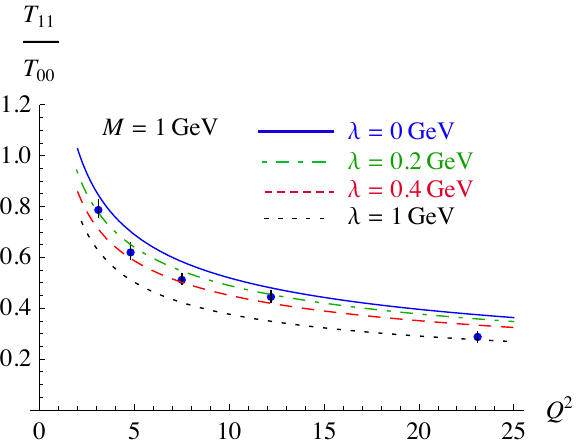}
\caption{Predictions for the ratio $T_{11}/T_{00}$ as a function of $Q^2$, compared  to the experimental data from H1.  Left: fixed $\lambda=0$, and various values for $M$. Right: fixed scale $M=1$ GeV, and various values of  $\lambda$.} 
	\label{T_11_00_param}
\end{figure}
 We refer the reader to publication \cite{Anikin:2011sa} for more details and now show some results. We have evaluated the ratios $T_{11}/T_{00}$ and $T_{01}/T_{00}$, which we compared with recent data from H1 \cite{Aaron:2009xp} and ZEUS \cite{Chekanov:2007zr}. In Figs. 1 and 2, we plot our results for the ratios $T_{11}/T_{00}$   $T_{01}/T_{00}$.   We get fairly good 
agreement with reasonable values of the two non-perturbative parameters $\lambda$ and $M$ involved in our model. Our calculation involves both Wandzura-Wilczek (WW) and genuine twist 3 contributions. It turns out that
with the input for coupling constants in DAs of $\rho-$meson, determined from the QCD sum rules, the WW contribution strongly dominates these two observables.

\begin{figure}[htbp]
	\hspace{1cm}
	\includegraphics[width=0.4\textwidth]{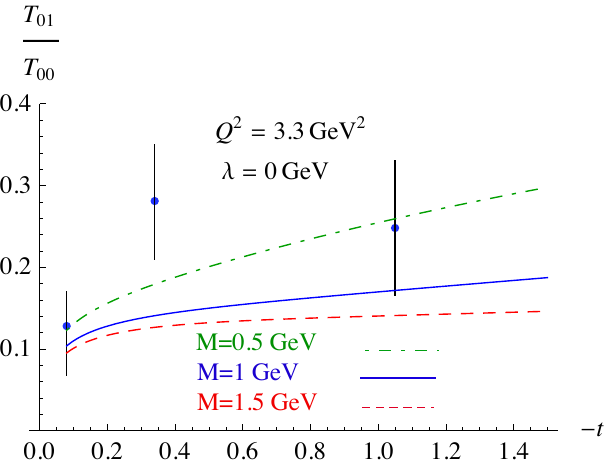}
\hspace{1cm}
\includegraphics[width=0.4\textwidth]{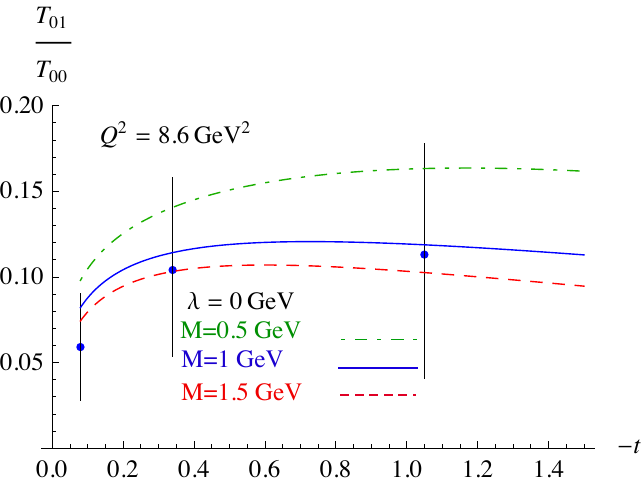}
	\caption{Predictions for the ratio $T_{01}/T_{00}$ as a function of $\left|t\right|$  for various values of $M$  compared with H1 data. Left:  $Q^2=3.3$ GeV$^2$. Right: $Q^2=8.6$ GeV$^2$. }
	\label{fig:T01delta_b}
\end{figure}

Other scattering amplitudes ratios have also been measured and  should be confronted to a $k_T$-factorization approach. The BFKL resummation effects have not been included. Although they are known  to be rather dramatic  at the level of scattering amplitudes,
we expect that for ratios of amplitudes considered here, they should be rather moderate.
Finally, let us stress that future electron-ion collider with high  energy and high luminosities \cite{Boer:2011fh}
will open  the opportunity  to study in details the electroproduction of meson.


\begin{thebibliography}{99}

 \bibitem{Anikin:2011sa}
  I.~V.~Anikin  {\it et al.},
  %``A phenomenological study of helicity amplitudes of high energy exclusive leptoproduction of the rho meson,''
 Phys. Rev. {\bf D84}, 054004 (2011) 
  [arXiv:1105.1761 [hep-ph]].
  %%CITATION = ARXIV:1105.1761;%%
  
    \bibitem{Anikin:2009}
I.~V.~Anikin  {\it et al.},
  %``QCD factorization of exclusive processes beyond leading twist: gamma*T ---> rho(T) impact factor with twist three accuracy,''
  Phys.\ Lett.\  {\bf B682}, 413-418 (2010) and Nucl.\ Phys.\  {\bf B828}, 1-68 (2010).
   %%CITATION = PHLTA,B682,413;%%
 %%CITATION = NUPHA,B828,1;%%


  
  \bibitem{Gunion:1976iy}
  J.~F.~Gunion, D.~E.~Soper,
  %``Quark Counting and Hadron Size Effects for Total Cross-Sections,''
  Phys.\ Rev.\  {\bf D15}, 2617-2621 (1977).
   %%CITATION =PHRVA,D15,2617;%%
  

  
  \bibitem{Navelet:1996jx}
  H.~Navelet, R.~B.~Peschanski, C.~Royon, S.~Wallon,
  %``Proton structure functions in the dipole picture of BFKL dynamics,''
  Phys.\ Lett.\  {\bf B385}, 357-364 (1996).
   %%CITATION = PHLTA,B385,357;%%
   
   \bibitem{Ginzburg:1985tp}
  I.~F.~Ginzburg, S.~L.~Panfil, V.~G.~Serbo,
  %``Possibility Of The Experimental Investigation Of The Qcd Pomeron In Semihard Processes At The Gamma Gamma Collisions,''
  Nucl.\ Phys.\  {\bf B284}, 685-705 (1987).
 %%CITATION = NUPHA,B284,685;%%


    
  \bibitem{Aaron:2009xp}
  F.~D.~Aaron {\it et al.} [ H1 Collaboration ],
  %``Diffractive Electroproduction of rho and phi Mesons at HERA,''
  JHEP {\bf 1005}, 032 (2010).
  %%CITATION = JHEPA,1005,032;%%
  
  \bibitem{Chekanov:2007zr}
  S.~Chekanov {\it et al.} [ ZEUS Collaboration ],
  %``Exclusive rho0 production in deep inelastic scattering at HERA,''
  PMC Phys.\  {\bf A1}, 6 (2007).
  [arXiv:0708.1478 [hep-ex]].
   %%CITATION = ARXIV:0708.1478;%%
  

\bibitem{Boer:2011fh}
  D.~Boer {\it et al.},
  %``Gluons and the quark sea at high energies: Distributions, polarization, tomography,''
   arXiv:1108.1713 [nucl-th].
 %%CITATION = ARXIV:1108.1713;%%
  


\end{thebibliography}
\end{document}